\documentclass[12pt,preprint]{aastex}
\shorttitle{NIR Variability of Sgr A*}

\begin{document}

\title{A near-IR variability study of the Galactic black hole: a red noise source with no detected periodicity}

\author{T. Do, A. M. Ghez, M. R. Morris, S. Yelda, L. Meyer, J. R. Lu}
\affil{Physics and Astronomy Department, University of California,
    Los Angeles, CA 90095-1547}
\author{S. D. Hornstein}
\affil{Center for Astrophysics and Space Astronomy, Department of Astrophysical and Planetary Sciences, University of Colorado, Boulder, CO 80309}
\author{K. Matthews}
\affil{California Institute of Technology, Pasadena, CA}

\email{tdo@astro.ucla.edu}

\begin{abstract}
We present the results of near-infrared (2 and 3 \micron) monitoring of Sgr A*-IR with 1 min time sampling using the natural and laser guide star adaptive optics (LGS AO) system at the Keck II telescope. Sgr A*-IR was observed continuously for up to three hours on each of seven nights, between 2005 July and 2007 August. Sgr A*-IR is detected at all times and is continuously variable, with a median observed 2 $\micron$ flux density of 0.192 mJy, corresponding to 16.3 magnitude at $K^{\prime}$. These observations allow us to investigate Nyquist sampled periods ranging from about 2 minutes to an hour. Using Monte Carlo simulations, we find that the variability of Sgr A* in this data set is consistent with models based on correlated noise with power spectra having frequency dependent power law slopes between 2.0 to 3.0, consistent with those reported for AGN light curves. Of particular interest are periods of $\sim 20$ min, corresponding to a quasi-periodic signal claimed based upon previous near-infrared observations and interpreted as the orbit of a `hot spot' at or near the last stable orbit of a spinning black hole. We find no significant periodicity at any time scale probed in these new observations for periodic signals. This study is sensitive to periodic signals with amplitudes greater than 20\% of the maximum amplitude of the underlying red noise component for light curves with duration greater than $\sim 2$ hours at a 98\% confidence limit.

\end{abstract}


\section{Introduction}

The existence of a super-massive black hole with a mass of $\sim 4\times10^6 M_\odot$ at the center of the Galaxy has now been firmly established from monitoring the orbits of the stars in the near-infrared (NIR) within 1 arcsecond of the location of the associated radio source Sgr A* \citep[e.g.][]{2002Natur.419..694S,2003ApJ...596.1015S,2003ApJ...586L.127G,2005ApJ...620..744G,Ghez2008}. Multi-wavelength detections of the radio point source at sub-millimeter, X-ray, and infrared wavelengths have also been made, showing that the luminosity associated with the black hole is many orders of magnitudes below that of active galactic nuclei (AGN) with comparable masses \citep{2001ARA&A..39..309M}. These observations have also shown that the emission from Sgr A* is variable \citep[e.g.,][]{2003ApJ...591..891B,2005ApJ...623L..25M,2005ApJ...628..246E,2007ApJ...667..900H,2007arXiv0712.2877M,2008A&A...479..625E}. Although it is now easily detected in its bright states when its flux increases by up to an order of magnitude over time scales of 1 to 3 hours, Sgr A* is difficult to detect in its faintest states at X-ray wavelengths because of the strong diffuse background, and in the near-infrared because of confusion with nearby stellar sources \citep{2003ApJ...591..891B,2007ApJ...667..900H}. Advances in adaptive optics (AO) technology have offered improved sensitivity to infrared emission from the location of Sgr A* against the stellar background, such that observations in its faint states are now possible \citep{2005ApJ...635.1087G}. Hereafter, the IR-luminous source Sgr A*-IR will be referred to simply as Sgr A*, recognizing that it is likely to be coincident with the radio source of that name. 

At both NIR and X-ray wavelengths, a possible quasi-periodic oscillation (QPO) signal with a $\sim20$ min period has been reported in light curves of Sgr A* \citep{2003Natur.425..934G,2004A&A...417...71A,2006A&A...455....1E}. Models that aim to produce QPO signals include both a class of models involving the Keplerian orbits of `hot spots' of plasma at the last stable orbit \citep{2006A&A...460...15M,2007MNRAS.375..764T} as well as models with rotational modulations of instabilities in the accretion flow \citep{2007ApJ...662L..15F}. Since the orbital period at the last stable orbit of a non-spinning black hole is 32 (M$_{bh}$/$4.2\times10^6M_\odot$) min, this putative periodic signal has been interpreted as evidence for a spinning black hole. The challenges for these claims are the relatively short time baselines of the observations (only a few times the claimed period), the low amplitude of the possible QPO activity, and the level of rigorous assessment of the statistical significance of the claimed periodicity. 

An alternative explanation for peaks in the periodograms seen in previous studies and interpreted as a periodic signal is that they are a sign of a frequency dependent physical process, commonly known as red noise \citep{1978ComAp...7..103P}. The power spectrum of such a physical process will display an inverse power law dependence on frequency, which manifests as light curves with large amplitude variations over long time scales and small amplitude variations over short time scales. The power spectrum of any individual realization of a red noise light curve will show statistical fluctuations around the intrinsic power law function, creating spurious peaks that can  lead to an interpretation of periodic activity. Variability studies of other accreting black hole systems like AGNs and Galactic X-ray binaries have shown that their power spectral densities are consistent with red noise. Several physical models have been proposed to produce the red noise light curves seen in AGNs \citep[e.g.][]{1997MNRAS.292..679L,2003MNRAS.341.1041A,2003MNRAS.339.1237V}; one common model that results in a red noise spectrum is from fluctuations in the physical parameters, such as the gas densities and accretion rate, at different radii of a turbulent magnetohydrodynamic accretion disk \citep{2001ApJ...560..659K}. While QPO signals have been unambiguously confirmed in X-ray binaries, no QPO signals in AGNs have been shown to be statistically different than red noise \citep{2001ApJ...562L.121B,2005A&A...431..391V}. Recent work by \citet{2008Bel} have also shown that a statistical analysis of X-ray light curves of Sgr A*, when including the contribution from red noise, show no indications of a QPO signal.

High sensitivity and high angular resolution near-infrared observations of Sgr A* have been obtained at the Keck II telescope utilizing new improvements in adaptive optics technologies to investigate the existence of a QPO signal as well as the timing properties of Sgr A*. These observations are described in Section \ref{sec:obs}. In order to test whether the variability of Sgr A* has the characteristics of red noise and to examine the possibility of a periodic signal, we have carried out a statistical analysis of the timing properties of the observed light curves that includes the possible contribution of red noise in the power spectrum to establish the significance of peaks in the periodograms. We find that the near-infrared variability of Sgr A* is entirely consistent with red noise, with no periodic signals detected on any night. Sections \ref{sec:per} details our analysis of the light curves. Lastly, in Section \ref{sec:dis}, we discuss the implications of our results for Keplerian models of the Sgr A* flux variability and compare the timing properties of Sgr A* with those of AGNs. 

\section{Observations and Data Reduction}
\label{sec:obs}

The Galactic center has been extensively imaged between 2005 and 2007 with the Keck II 10 m telescope using the natural and laser guide star adaptive optics (NGS and LGS AO) \citep{2006PASP..118..297W, 2006PASP..118..310V} system and the NIRC2 near infrared camera (P.I. K. Matthews). For this study we include all nights of LGS-AO observations at $K^\prime$ (2 \micron) that had sampling of 1-3 minutes, a total time baseline of at least $\sim1$ hour, and at least 60 data points. We also include one night of NGS observations at $L^\prime$ that satisfied the same criteria. As summarized in Table \ref{table:obs}, this resulted in a selection of 7 data sets with durations ranging from 80 min to 3 hours. 

A detailed description of LGS AO observations of the Galactic center are described in \citet{2005ApJ...635.1087G}; here, we only summarize the setup for our observations. The laser guide star was propagated at the center of our field and for low order tip-tilt corrections, we used the $R = 13.7$ mag star, USNO 0600-28577051, which is located $\sim19\arcsec$ from Sgr A*. Most of the images were obtained using the $K^{\prime}$ band-pass filter ($\lambda_o = 2.12$ \micron, $\Delta\lambda = 0.3$ \micron) and were composed of 10 coadded 2.8 sec exposures, for a total integration time of 28 sec. The remaining set of observations from 2005 July 28 was taken through the $L^{\prime}$ band-pass filter ($\lambda_o = 3.78$ \micron, $\Delta\lambda = 0.7$ \micron). For five of the $K^{\prime}$ nights, the time interval between each image is about 50 seconds, with dithers every three minutes. $K^{\prime}$ images from 2006 July 17 were sampled at 3 minute intervals but were not dithered. The $L^{\prime}$ observations had one minute sampling, and were also not dithered. The 3 min dithers affect the timing analysis by the presense of a spike at that frequency in the periodograms (this is well reproduced by Monte Carlo simulations of the effects of sampling). 

\begin{figure}
\centering
\includegraphics[width=3in,angle=90]{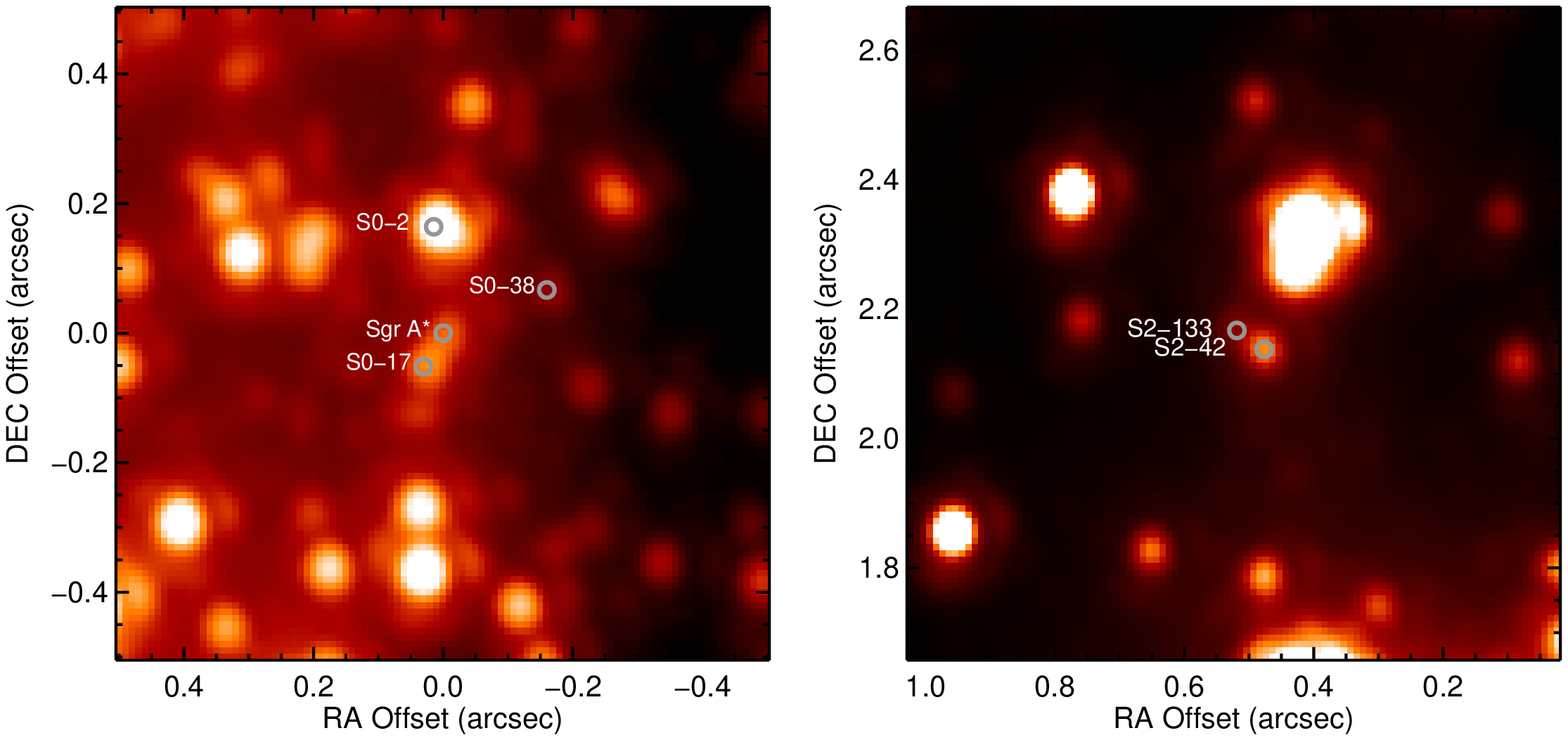}
\caption{\textbf{Left}: a $K^\prime$ image from 2006 May 3 of the central 0.\arcsec5 around Sgr A* with a logarithmic intensity scale so that the faint sources can be more easily seen. Sgr A* (K$^\prime = 15.8$ in this image) is in the center of the image along with the comparison stars, S0-2, S0-17 and S0-38. The image is oriented with north up and east to the left, with offsets in projected distance from Sgr A*. \textbf{Right}: image from the same night of the pair of comparison sources, S2-42 ($K^\prime =$ 15.5) and S2-133 ($K^\prime =$ 16.7), with a flux ratio similar to that of S0-17 and Sgr A* when Sgr A* is faint. These two stars also has a similar separation in the plane of the sky as S0-17 and Sgr A* ($\sim$50 mas).}
\label{fig:img}
\end{figure}

Photometry was performed on the individual images using the point spread function (PSF) fitting program \textit{StarFinder} \citep{2000A&AS..147..335D}. The program was enhanced as described in \citet{2007ApJ...667..900H} to include the \textit{a priori} knowledge of the location of the near-infrared position of Sgr A* and nearby sources in order to facilitate the detection of Sgr A* at faint flux levels in these short exposures. To do this, for each night of observation, the position of Sgr A* and nearby sources was determined in a nightly-averaged image produced by a weighted average of individual images from that night (see Figure \ref{fig:img}). We then used the knowledge of the location of all the sources as fixed inputs into \textit{StarFinder} to more accurately fit for the flux contribution of sources near Sgr A* in the individual short exposure images. To compensate for seeing changes through the night, a different PSF was constructed for each image. We also include only images with Strehl ratios greater than 20\% to minimize large errors in the photometry from bad seeing conditions, which resulted in dropping only about 10 data points out of all nights. On average, the Strehl ratio at $K^{\prime}$ was 32\%, with the full width of the core at half-maximum intensity (FWHM) $\sim60$ mas as measured from the relatively isolated star IRS 33N.  We are able to detect Sgr A* at all times, even at its faintest flux levels. The gaps in the data are from technical disruptions in the observations. 

\begin{figure}
\centering
\includegraphics[width=3in,angle=90]{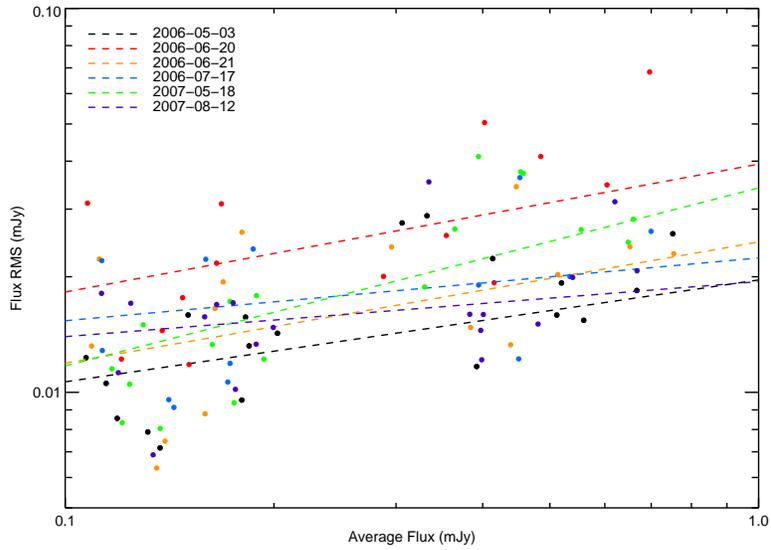}
\caption{Plot of power law fits to the rms fluxes for stars within 0.\arcsec5 of Sgr A* with brightnesses within the range of observed brightness variations in the $K^\prime$ light curves from Sgr A*. The photometric noise properties are similar for each night.}
\label{fig:noisestats}
\end{figure}

Photometric calibrations were performed relative to the list of non-variable sources from \citet{2007ApJ...659.1241R} at $K^\prime$ and IRS 16C ($L^{\prime} = 8.14$ mag) and IRS 16NW ($L^\prime = 8.43$ mag) at $L^\prime$ \citep{1996ApJ...470..864B}. The photometric error at each flux density level seen in Sgr A* was estimated by fitting a power law to the rms uncertainty in the flux for all non-variable stars in the same range of brightnesses observed for Sgr A* within 0.\arcsec5 of the black hole (see Figure \ref{fig:noisestats}). We find the typical dependence of the photometric error, $\sigma$, on flux density, F, to be: $\sigma \approx 0.2 F^{0.3}$ mJy. The flux measurement uncertainties are comparable for all nights except 2006 June 20, the night with the worst seeing. Within the range of observed Sgr A* fluxes, we are on average able to achieve between 3 to 15\% relative photometric precision for each 28 sec $K^\prime$ exposure. A source of systematic error in the flux measurements is the proximity of Sgr A* to unresolved sources, which could contribute flux. This contribution is only likely to have an impact when Sgr A* is faint \citep{2007ApJ...667..900H}, but for the purpose of this variability study, this effect is likely only a systematic offset in the mean flux density and a source of white noise (for more details, see Section \ref{sec:flux}). For comparison to Sgr A*, we also analyze the light curves of the nearby stars S0-17 and S0-38 at $K^\prime$ and S0-2 at $L^\prime$. S0-17 ($K^\prime = 15.5$ mag) was chosen because it is spatially closest to Sgr A*, with a projected distance from Sgr A* of $\sim 56$ mas in 2006 May to $\sim 48$ mas in 2007 August; monitoring S0-17 is helpful to ensure that the variations in flux seen in Sgr A* are not a systematic effect of seeing or bias from nearby sources. The star S0-38 ($\sim 0.11$ mJy, $K^\prime \sim 17$ mag), $\sim 0\arcsec.2$ from Sgr A*, was chosen as a stellar reference because it has a similar flux to the faintest observed emission and given its proximity to Sgr A*, its photometry will be affect similarly from the unresolved stellar background. Figure \ref{fig:img} shows an image of this region and the location of the comparison sources with respect to Sgr A*.

In order to characterize possible effects on the photometry of Sgr A* by S0-17, we also use the photometry of two stars with separations and flux ratios similar to that of S0-17 and Sgr A* when Sgr A* is faint. The two stars, S2-42 (K$^\prime$=15.5) and S2-133 (K$^\prime$=16.7), are located about $2\arcsec$ from Sgr A* and are separated by $\sim50$ mas (Figure \ref{fig:img}). The rms variability of S2-42 and S1-133 is about 5\% and 15\%, respectively, similar to the photometric precision we would have predicted based upon our power law fits to the rms stability of stars near Sgr A*. Thus, we can be confident that the photometry of Sgr A* at its faintest is similar to stars of that magnitude despite the proximity of S0-17. 

\section{Results and Analysis}

\subsection{Flux Distribution}
\label{sec:flux}

\begin{figure}
\centering
\includegraphics[width=4.5in,angle=90]{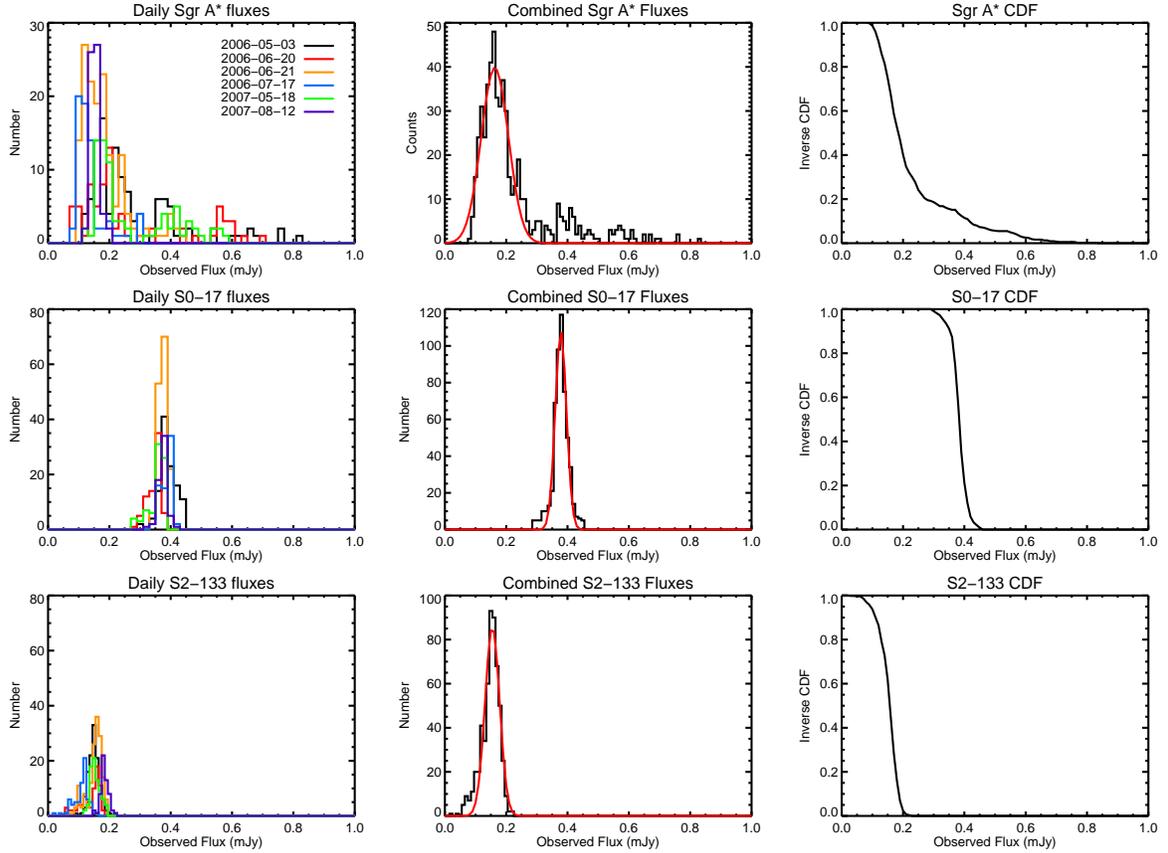}
\caption{\textbf{Top}: Sgr A* flux distribution for each of the 6 $K^{\prime}$ nights (left), the combined fluxes from all nights (middle), and the inverse cumulative distribution function for the combined fluxes (right). \textbf{Middle}: The corresponding plots for S0-17, the non-variable comparison source used in this study. The total histogram of fluxes observed from S0-17 is consistent with a single Gaussian. \textbf{Bottom}: the flux distribution for S2-133, a $K^\prime = 16.7$ magnitude star for comparison with the faint states of Sgr A*.}
\label{fig:hist}
\end{figure}

In order to characterize the range of fluxes observed from Sgr A*, we have constructed histograms of fluxes from each night as well as the combined histogram from all nights (Figure \ref{fig:hist}). Unless otherwise stated, the fluxes in this paper are observed fluxes and not corrected for extinction to Sgr A*. Where indicated, de-reddened fluxes have been calculated by assuming $A_v = 30$ \citep{2001A&A...366..106M} and extinction law $A_{K^\prime} = 0.1108 A_v$ \citep{1985ApJ...288..618R}. The comparison sources do not appear to be variable on the time scales probed in this study and have fluctuations consistent with Poisson noise. The flux distribution for the star S0-17 is consistent with a Gaussian centered at 0.37 mJy at $K^{\prime}$, with a standard deviation of 0.02 mJy; this suggests that we are able to reproduce the flux of S0-17 at the 5\% level between different nights. A bias in the photometry from the proximity of S0-17 to the light curve from Sgr A* should manifest itself as a  difference in the mean flux of S0-17 between 2006 and 2007 because S0-17 moved closer to Sgr A* in the plane of the sky between the two years. The fact that we observe the same mean flux from S0-17 between different years is also confirmation that there is little bias in the photometry of either S0-17 or Sgr A*. The light curves of S0-17 are also stable on each night, independent of the flux of Sgr A* and shows no greater variance than other stars of the same brightness in the region (within 0.\arcsec5 of Sgr A*), showing that PSF fitting from \textit{StarFinder} is able to able to properly account for the flux of both sources. 

\begin{figure}
\centering
\includegraphics[width=5in]{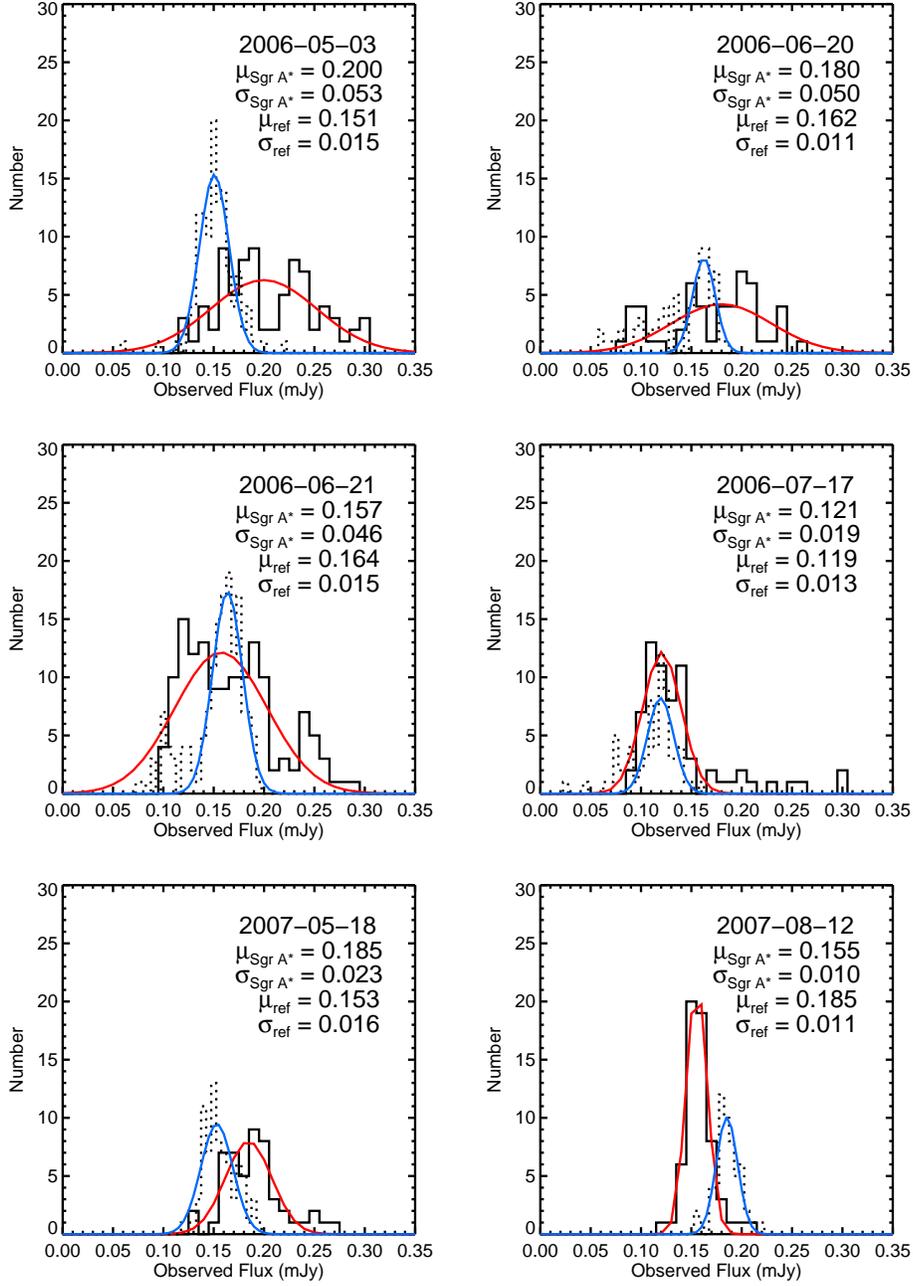}
\caption{The flux distribution of Sgr A* (solid) of fluxes less than 0.3 mJy and the reference source, S2-133 (dotted) for each night along with the best fit Gaussian. On four of the six $K^\prime$ nights, the width of the Sgr A* flux distribution is greater than the reference, while in the remaining two nights, the widths are comparable. Labels are units of mJy.}
\label{fig:nightly_hist}
\end{figure}

The cumulative distribution function for Sgr A* shows that it is brighter than S0-17 about 15\% of the time at $K^{\prime}$. The median flux density of Sgr A* is at 0.192 mJy ($K^{\prime} = 16.3$ mag), or a de-reddened flux of 4.10 mJy. The flux histogram for Sgr A* is not well fitted by a Gaussian because it has a long tail in the distribution of flux densities at high flux densities. However, if the tail of the distribution is excluded, the flux distribution below 0.3 mJy, is well fit by a Gaussian with a mean of 0.158 mJy and a standard deviation of 0.05 mJy. The latter is larger than nearby sources with comparable flux densities, indicating that Sgr A* is intrinsically variable; for example, the flux distribution for S2-133 has a FWHM of $\sim0.02$ mJy (see Figure \ref{fig:nightly_hist}). 

\begin{figure}
\centering
\includegraphics[width=3in,angle=90]{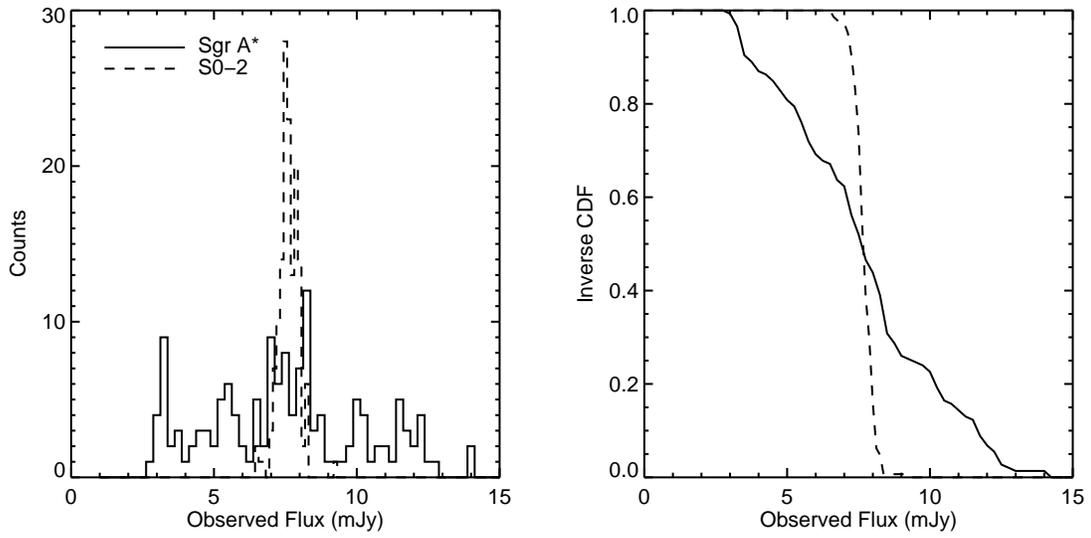}
\caption{\textbf{Left} - Sgr A* and S0-2 flux distributions on the night of 2005 July 28 at $L^{\prime}$ . \textbf{Right} - the corresponding cumulative distribution function for Sgr A* and S0-2. Both CDFs are consistent with Gaussian distributions. }
\label{fig:hist_lp}
\end{figure}

The flux density distribution of Sgr A* for the $L^\prime$ night shows a much larger width than the comparison star S0-2 (Figure \ref{fig:hist_lp}). The distribution appears to be symmetric about the mean and the CDF is consistent with a Gaussian with mean flux density of 7.69 mJy with $\sigma = 2.81$ mJy, compared to S0-2 with a mean flux density of 7.67 mJy with $\sigma = 0.37$ mJy.

While the bright flux levels ($>$ 0.3 mJy) can be unambiguously attributed to the black hole, the source of emission from the location of Sgr A* when it is faint is less certain. The region immediately around Sgr A* is also the location of peak stellar density, which raises the possibility of flux contamination from either an unresolved population of stars or a very faint star in a close orbit at that location. We find that the faintest observed flux density at $K^\prime$ is $0.082\pm0.017$ mJy ($K^\prime = 17.2$) or $1.75\pm$0.36 mJy de-reddened from the location of Sgr A* (consistent with limits observed by \citet{2002ApJ...577L...9H}). Comparison of the flux distribution of S2-133 with that of Sgr A* below 0.3 mJy shows that, for four of the six $K^\prime$ nights, Sgr A* has a larger variance than S2-133 (see figure \ref{fig:nightly_hist}), indicating that Sgr A* is more variable than expected for a stellar source even at the faintest levels. On the remaining two nights, the variance of Sgr A* is similar to that of S2-133. On one of these two nights, 2007 August 12, Sgr A* was fainter than 0.22 mJy for the entire duration of our observation, which makes this night ideal for timing analysis of Sgr A* at its faintest flux density levels. Though the flux distribution looks similar to a star, the structure function and the periodogram shows a slightly steeper slope than expected for Gaussian noise and as compared to the stellar stellar comparison sources (see Sections \ref{sec:per} and \ref{sec:sf}). Furthermore, the $K^\prime - L^\prime$ color for 2006 July 17, previously reported in \citet{2007ApJ...667..900H}, is constant and significantly redder, even at its faintest on that night, than from a stellar source. At the faintest fluxes, between 0.10 and 0.15 mJy, the mean $K^\prime - L^\prime$ spectral slope, corrected for extinction, of Sgr A* has an average  power law exponent of -0.17$\pm0.32$, compared to a slope of $-0.6\pm0.2$ from \citet{2007ApJ...667..900H}. We estimate that a stellar source, which would have de-reddened spectral slope of 2, could contribute a maximum of about 35\% of the flux to account for the difference in the spectral slope. This leads us to conclude that, even when the emission is faint, a large fraction of the flux arising from the location of Sgr A* is likely non-stellar and can be attributed to physical processes associated with the black hole. 

\subsection{Light Curves and Timing Analysis}

\begin{figure}
\centering
\includegraphics[width=4.5in,angle=90]{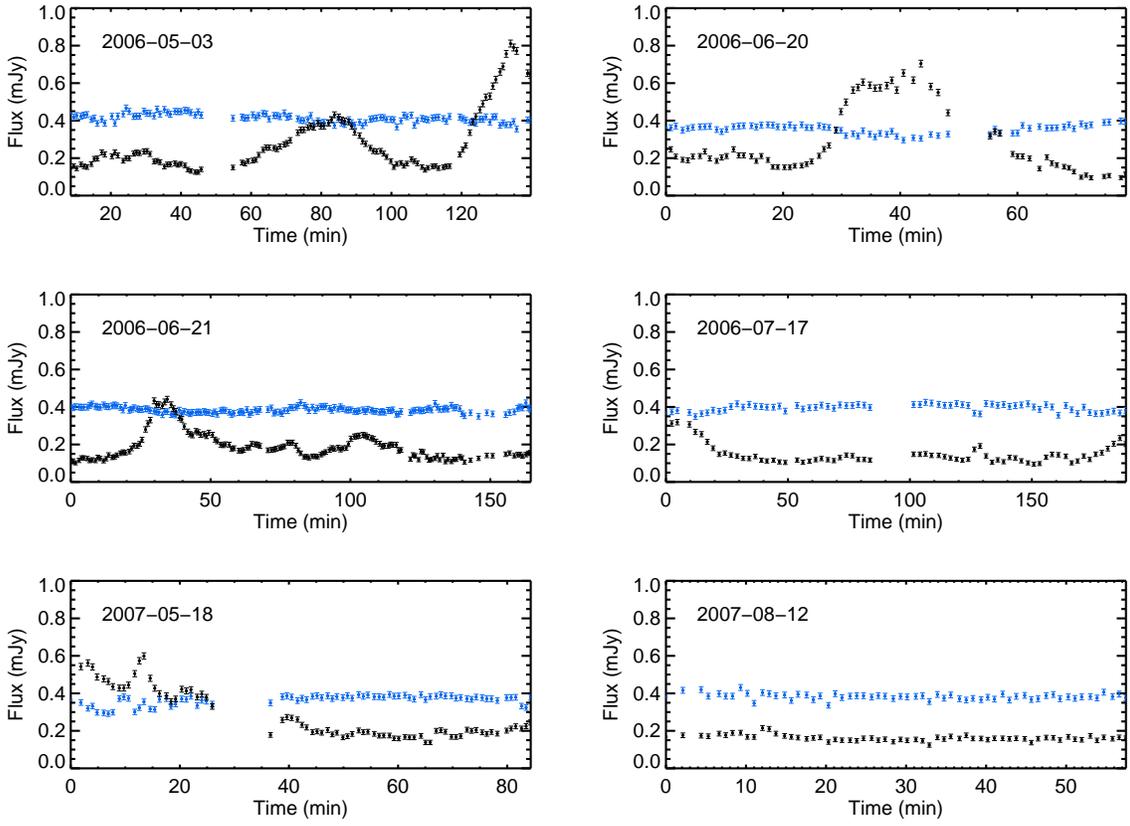}
\caption{Sgr A* light curves (black) at $K^{\prime}$ with the star S0-17 (blue) for comparison.}
\label{fig:sgralc}
\end{figure}

\begin{figure}
\centering
\includegraphics[width=5in]{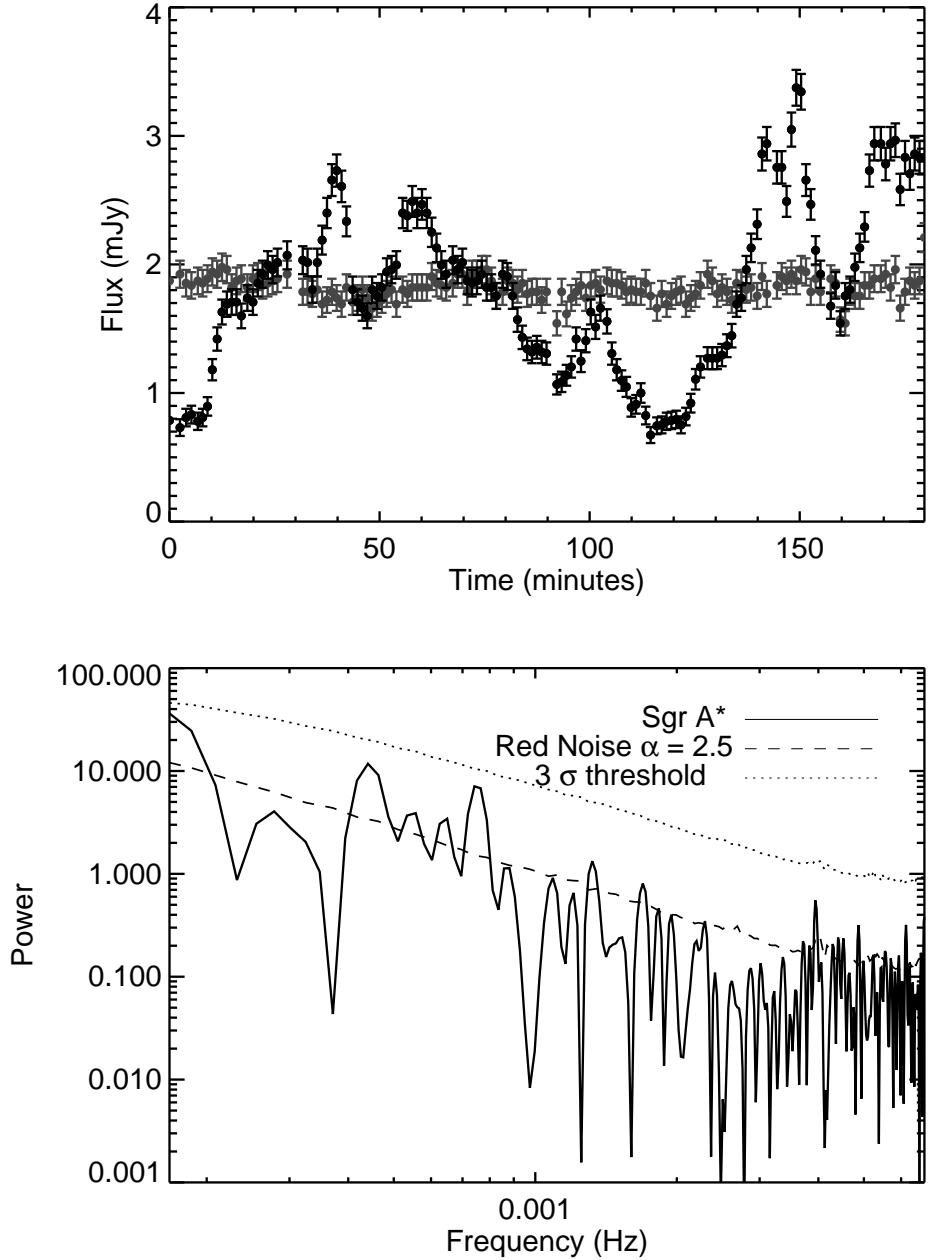}
\caption{\textbf{Top:} The light curve of Sgr A* (black) and comparison source S0-2 (grey) at $L^{\prime}$ on 2005 July 28. \textbf{Bottom:} Normalized Lomb-Scargle periodograms of  Sgr A* on this night (black). Also plotted is the 3 $\sigma$ significance threshold determined from Monte Carlo simulations of red noise with a power law index, $\alpha = 2.5$ (dotted). The average of 10$^5$ periodograms with $\alpha = 2.5$ at the same time sampling is also shown for comparison (dashed line).}
\label{fig:sgralc_lp}
\end{figure}

Figures \ref{fig:sgralc} and \ref{fig:sgralc_lp} shows the resulting light curves for Sgr A* and a non-variable comparison source for each night of observation. While comparison sources show no significant time variable emission, Sgr A* shows variations on time scales ranging from minutes to hours, with peak emission that can be 10 times higher than during its faintest states. The emission peaks are time symmetric, with similar rise and fall times. 

In order to characterize the variability of Sgr A*, we have carried out the following three different approaches to timing analysis: (1) periodograms (2) structure functions and (3) auto-correlations. The periodogram analysis, presented in section \ref{sec:per}, is effective at pulling out periodic structure in light curves and therefore is optimal for assessing the presence of any periodicities, such as the proposed $\sim 20$ min QPO reported in previous experiments \citep[e.g.,][]{2003Natur.425..934G,2006A&A...455....1E}. Both the periodogram and the structure function can also be used to measure the underlying power spectral density (PSD), which can then be used to explore similarities to the variability observed in AGNs. We also compute the auto-correlation for each night to look for possible differences in the variability at each time scale between each night. 

\label{sec:lightcurves}

\subsubsection{Periodogram}
\label{sec:per}

It is important to consider all possible sources of noise when testing for periodicity in light curves. While peaks in the periodograms are a good place to start searching for periodicity, the peaks must have significantly more power than those produced by non-periodic processes to be unambiguously attributed to a true periodicity in any variable source. White (Gaussian) noise processes are unlikely to lead to large peaks in the periodograms because they contribute equal power at all frequencies. However, time-correlated physical processes can result in variability that is frequency dependent. One common variability characteristic - often seen in AGN light curves - is red noise, which can lead to spurious signals in a power spectrum or periodogram from a data set having a time baseline only a few times longer than that of the putative period, since it will show large amplitude fluctuations at low frequencies and small amplitudes at high frequencies. This can lead to relatively large stochastic peaks in the power spectrum at low frequencies, far above what would be expected from white noise. We emphasize that, although the term for this type of power law dependence of the flux variability is `red noise', this variability arises from physical processes from the source and is not a result of measurement uncertainties such as Poisson noise, which behaves like white noise in its power spectrum.

One of the goals in this timing analysis is to test whether a purely red noise model can explain the variability of Sgr A*. The PSD of a red noise light curve is a power law, with greater power at lower frequencies: $P(f) \equiv f^{-\alpha}$, where $f$ is the frequency and $\alpha$ is the power law index. For example, $\alpha = 0$ for white noise and $\alpha = 1$ for classical flicker noise \citep{1978ComAp...7..103P}. All red noise simulations in this paper were produced by an algorithm detailed in \citet{1995A&A...300..707T}, which randomizes both phase and amplitude of an underlying power law spectrum and then inverse Fourier transforms it into the time domain to create light curves. Our procedure for producing simulated light curves is as follows: (1) a light curve is produced from a PSD with a specific power law slope evenly sampled at half the shortest observed time sampling interval, with a duration at least 10 times as long as the observed light curve (rounded up to the nearest power of 2 for computational efficiency of the fast Fourier transform). This length was chosen based upon the suggestion by \citet{2002MNRAS.332..231U} to avoid a 'red noise leak' where power is distributed from frequencies lower than that sampled by the observation into observed frequencies. We find aliasing to be a negligible effect, because the light curves are generated at higher temporal resolution than the observations. (2) this light curve is then split into 10 non-overlapping segments to reduce simulation time \citep{2002MNRAS.332..231U}. Each segment is then re-sampled at the exact sampling times used during the specific night that we are simulating. (3) since the simulation has an arbitrary flux scale, we scale the light curves to have the same mean flux level and standard deviation as that night. We also include the effects of measurement noise in the simulations by adding Gaussianly distributed noise to each simulated data point. We use our measurements of the photometric error as a function of flux densities (Section \ref{sec:obs}) of non-variable stars within 0\arcsec.5 of Sgr A* to account for the flux density dependence in the noise for each simulated data point. 

Instead of computing the PSD, which is often used for evenly sampled data, we searched for periodicity by computing a related function for unevenly sampled data: the normalized Lomb-Scargle periodogram \citep{1989ApJ...338..277P}; given a set of data values $h_{i}$, $i = 1, \ldots, N$ at times $t_{i}$ the periodogram is defined as:

\begin{equation}
P_{N}(\omega) \equiv \frac{1}{2\sigma^{2}}\left\{\frac{[\sum_{j}(h_{j}-\overline{h})\cos{\omega(t_{j}-\tau)}]^{2}}{\sum_{j}\cos{^{2}\omega(t_{j}-\tau)}} + \frac{[\sum_{j}(h_{j}-\overline{h})\sin{\omega(t_{j}-\tau)}]^{2}}{\sum_{j}\sin{^{2}\omega(t_{j}-\tau)}}\right\}
\end{equation}
where $\omega$ is the angular search frequency, $\overline{h}$ and $\sigma^{2}$ are the mean and variance of the data respectively. The constant $\tau$ is an offset introduced to keep the periodogram phase invariant: 
\begin{equation}
\tan{(2\omega \tau)} = \frac{\sum_{j}\sin{2\omega t_{j}}}{\sum_{j}\cos{2\omega t_{j}}}
\end{equation}
Since the periodogram is normalized by the variance of the flux, a light curve consisting of only white noise, or equivalently, red noise with a power law $\alpha = 0$, will have an average power of 1 at all frequencies.

\begin{figure}
\centering
\includegraphics[width=4.5in,angle=90]{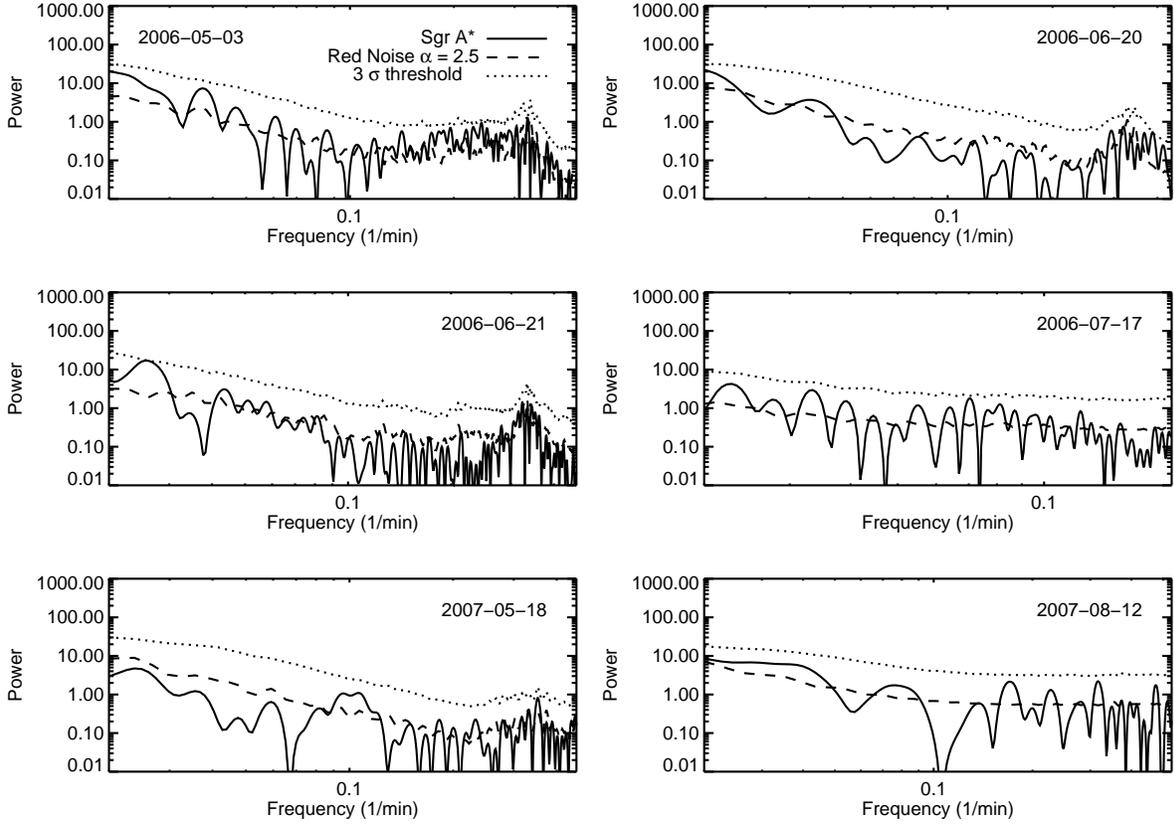}
\caption{Normalized Lomb-Scargle periodograms of the Sgr A* light curves (black) at $K^{\prime}$. Also plotted is the 3 $\sigma$ significance threshold determined from Monte Carlo simulations of red noise with a power law index, $\alpha = 2.5$ (dotted line). The average of 10$^5$ periodograms with $\alpha = 2.5$ at the same time sampling is also shown for comparison (dashed line).}
\label{fig:sgraPer}
\end{figure}

\begin{figure}
\centering
\includegraphics[width=5in]{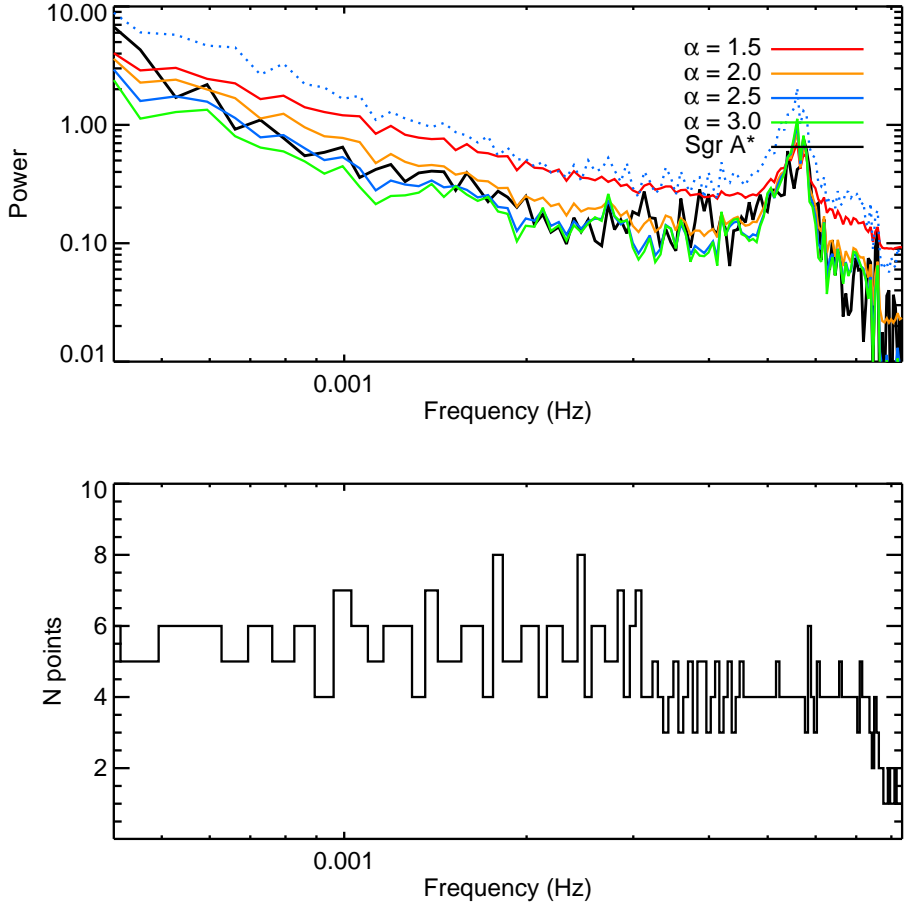}
\caption{\textbf{Top}: averaged periodogram from 5 nights of $K^{\prime}$ observations (black line) with durations greater than 80 min (resulting in the exclusion of 2007 Aug 12). Colored lines are the average of $10^{4}$ Monte Carlo simulations of combining data sets with the same sampling, for three different red noise power laws. The dotted blue line corresponds to the 3 $\sigma$ threshold of power for red noise with $\alpha = 2.5$. The large spike at 0.0056 Hz (3 min) is an artifact resulting from the regular dithers in the observations. \textbf{Bottom}: plot of the number of points contributing to the averaged periodogram at each frequency.}
\label{fig:perCombo}
\end{figure}

The normalized Lomb-Scargle periodogram was computed for each light curve, oversampled by a factor of 4 times the independent Fourier intervals in order to increase the sensitivity to periods between the Fourier frequencies (Figures \ref{fig:sgralc_lp} and \ref{fig:sgraPer}). Assuming that the physical source of the variability is stationary, we averaged together the periodograms for the five $K^\prime$ nights which have durations longer than 80 minutes (Figure \ref{fig:perCombo}). The combined periodogram excludes the 2007 August 12 night because it is less than an hour long, leading to poor sampling at low frequencies compared to the other nights. We combined the periodograms by averaging the Lomb-Scargle power at linearly space frequency bins. The combined periodogram is consistent with red noise, except for the peak corresponding to the time scale of the three minute dithers. To characterize the underlying spectrum, we have performed Monte Carlo simulations combining red noise light curves with the same sampling as the data set. We tested several different underlying PSD and found that the combined periodogram is consistent with power law indices between 2.0 and 3.0, with no periodic components. This model is able to reproduce the slope of the periodogram, the increase in power at three minutes from dithering, and the flattening of the periodogram at very low frequencies caused by poor sampling at those frequencies. The 3 min peak in the periodogram is repoduced very well by the simulations, showing that the simulations are correctly accounting for the effects of sampling. Figure \ref{fig:perCombo} shows the results of Monte Carlo simulations with power law indices 1.5, 2.0, 2.5, and 3.0. The simulations shows that the resulting periodograms tend to be flatter than the intrinsic PSD because the limited time sampling at low frequencies results in poor sensitivity to long time scale variations characteristic of steeper power laws. Because the periodogram sampling is poor at frequencies below 40 min we will ignore those frequencies in all subsequent analysis. We find the resulting periodograms show less variation for intrinsic PSD $\alpha > 2$, which suggests that we have a better constraint on the lower limit than on the upper limit of our estimate for the slope of the PSD. While red noise models with values of $\alpha$ between 2.0 to 3.0 appear to be consistent with the average periodogram, we will use $\alpha = 2.5$ for the simulations in this work to provide a baseline for comparison. More light curves will be necessary to determine a reliable intrinsic PSD of Sgr A* (if it does not vary between nights).  Where appropriate, we have also run simulations with a range of $\alpha$ values to investigate its effects on the statistical significance. 

Although there is no evidence for QPO activity in the combined periodogram, we also test the case of a transient QPO phenomenon in each night. We therefore used our best fit of $\alpha = 2.5$ for the power law of the combined periodogram to test each night for statistically significant deviations from a purely red noise model. Our criterion for a statistically significant peak in the periodogram is that its power must be above the 99.7\% (3$\sigma$) confidence interval from a Monte Carlo simulation with $10^5$ realizations of red noise light curves. This method of establishing the significance of peaks in the periodogram is similar to the one proposed by \citet{2005A&A...431..391V} for evenly sampled data, but modified here to account for our unevenly sampled data set by using the Lomb-Scargle periodogram instead of the Fourier transform. The resulting periodograms are shown in Figure \ref{fig:sgraPer}. The individual periodograms show peaks at low frequencies, but these appear to be consistent with red noise, with no peaks having power greater than the 3$\sigma$ threshold derived from the Monte Carlo simulations. No significant peaks were found when we repeated the same procedure for PSD power law indices between 1.0 and 3.0.

\begin{figure}
\centering
\includegraphics[width=4.5in]{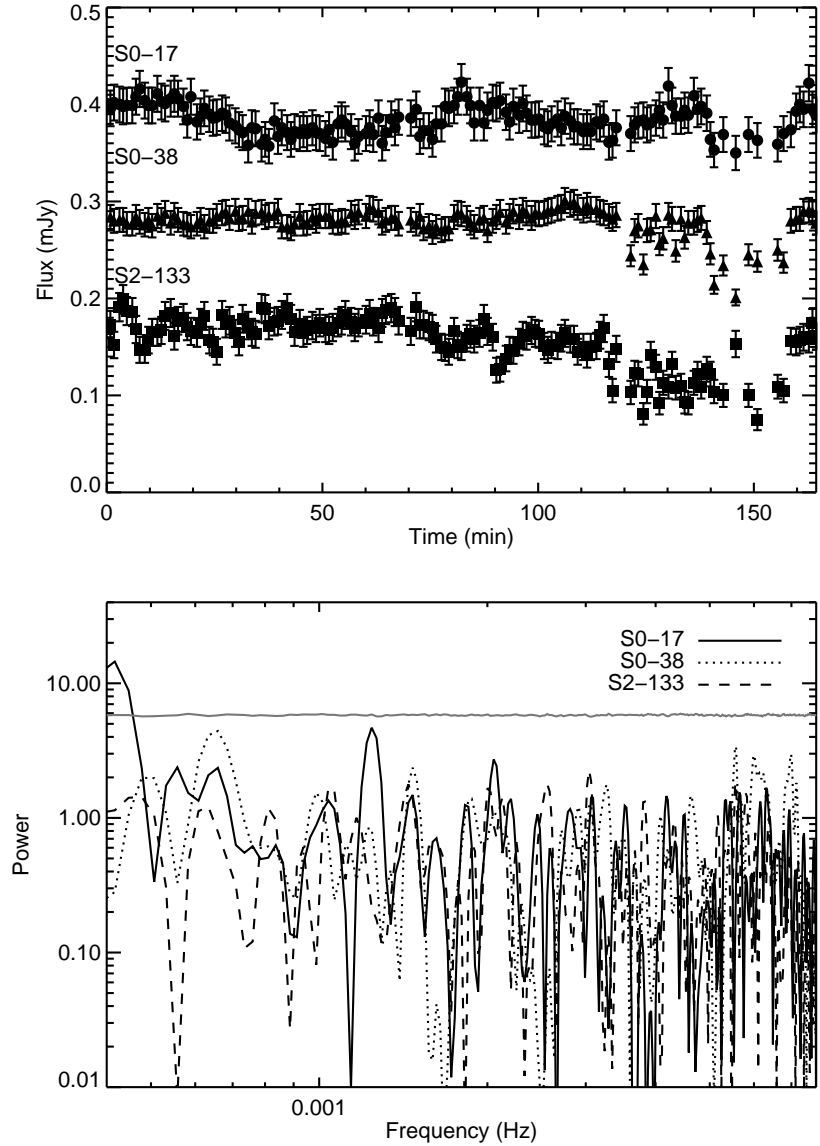}
\caption{\textbf{Top}: Light curves of the stellar comparison sources from 2006-06-21. The light curve of S0-38 is offset by 0.18 mJy for clarity. \textbf{Bottom:} The corresponding periodograms of the comparison sources. In solid gray is the 3 $\sigma$ threshold for white noise ($\alpha = 0$).}
\label{fig:refPer}
\end{figure}

The periodograms of the comparison sources are much flatter than that of Sgr A*, as expected for white noise sources. However, there is some power at very low frequencies ($<$ 40 min), which may be the result of some small time correlated systematic in the observation such as variations in seeing or AO performance over a night. Light curves and periodograms from 2006-06-21 of the three comparison sources S0-17, S0-38, and S2-133 are shown in figure \ref{fig:refPer}. The periodograms of these sources from the other nights of observations are also comparably flat.

\begin{figure}
\centering
\includegraphics[width=4.5in]{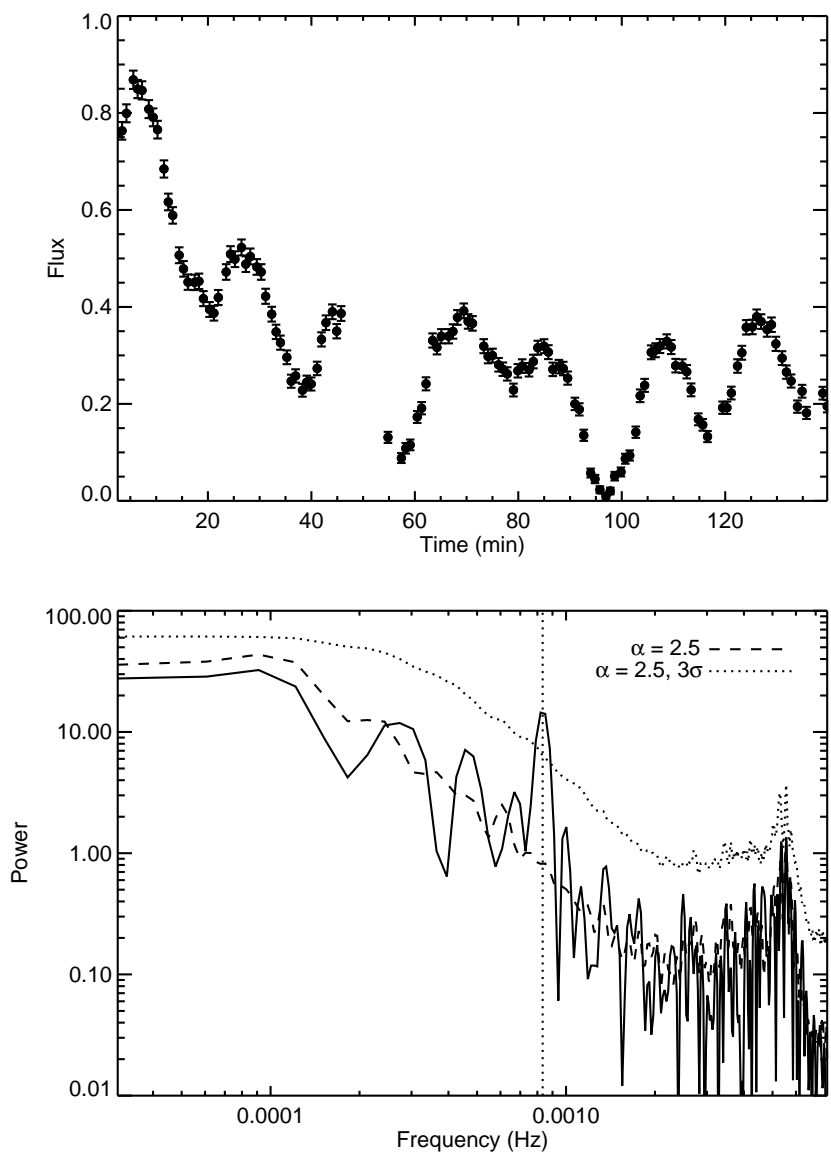}
\caption{\textbf{Top:} Simulated light curve with the same time sampling as 2006 May 03 with a 20 minute periodic signal with an amplitude 15\% of the maximum flux seen that night. Error bars in the light curve are based upon the noise properties of that night. \textbf{Bottom:} The corresponding periodogram of the simulated light curve (solid line) along with the average (dashed) and $3 \sigma$ threshold of red noise power from the Monte Carlo simulations (dotted line). We can recover the signal at 20 min (vertical dotted line) at above our established confidence threshold.}
\label{fig:sim_qpo}
\end{figure}

By adding an artificial sinusoidal periodic signal to the red noise simulations, we can address our sensitivity for detecting QPOs in the presence of red noise. In order to test this, a periodic sinusoidal signal was added to a simulated light curve with a red noise PSD slope of 2.5, with an amplitude that is a fraction of the maximum signal in the underlying red noise light curve. This fractional amplitude of the periodic signal was increased until the periodogram showed a peak at that frequency above the previously determined 3$\sigma$ threshold over 98\% of the time. For example, for the time sampling and duration corresponding to the observations on 2006 May 3, we find that we are able to detect a 20 min periodic signal with an amplitude that is at least 20\% of the maximum flux density (Figure \ref{fig:sim_qpo}). The sensitivity for the detection of a QPO increases with frequency because the underlying red noise component has less power at higher frequencies (e.g. a 10 (40)-minute QPO signal will result in power greater than expected from red noise when its amplitude is 5\% (30\%) of the maximum flux density). Our sensitivity for the detection of a periodic signal is similar for the other observed light curves with similar durations ($>$ 2 hours). 

\begin{figure}
\centering
\includegraphics[width=3in,angle=90]{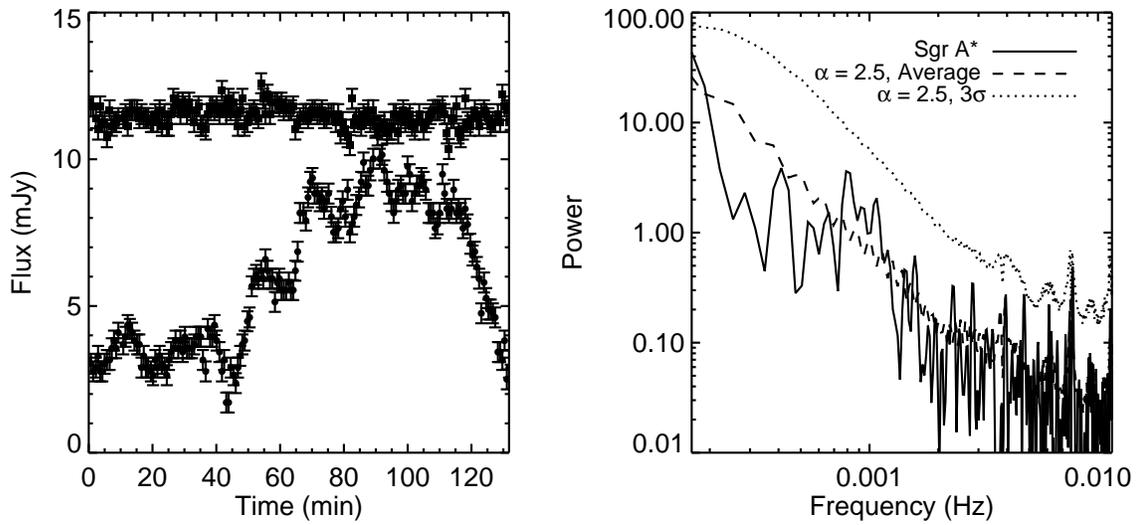}
\caption{\textbf{Left:} The de-reddened light curve of Sgr A*, from 2003 June 16, as extracted by \citet{2006A&A...460...15M} (filled circles) along with a comparison source, called S1 in that paper (squares, offset by 6 mJy for clarity). \textbf{Right:} The periodogram of the Sgr A* light curve (solid) with the mean power (dashed) and 3$\sigma$ thresholds set by Monte Carlo simulations of red noise with a power law slope of $\alpha = 2.5$ (dotted).}
\label{fig:03jun}
\end{figure}

We note that attempting to remove the broad, 40-100 minute emission peaks with a low-order functional fit to increase sensitivity to periodic signals will introduce a statistical bias, because that will only remove some combination of low frequency power without actually removing the entire red noise component. Because the light curves are consistent with a red noise process at all time scales, including the longer broad maxima, any statistical analysis must be performed without first modifying the light curves. For example, the light curve reported by \citet{2003Natur.425..934G}, the first used to argue for the possible presence of a periodic variation, shows no significant periodicity when subjected to the red noise analysis described here. Figure \ref{fig:03jun} shows the analysis of this night using the light curve re-extracted by \citet{2006A&A...460...15M}. A significant signal was found when a low order polynomial fit to the light curve was removed in \citet{2006A&A...460...15M}. However, this significance may be biased by the removal of a number of low frequency components from the light curve.

\subsubsection{Structure Function}
\label{sec:sf}

\begin{figure}
\centering
\includegraphics[width=3in,angle=90]{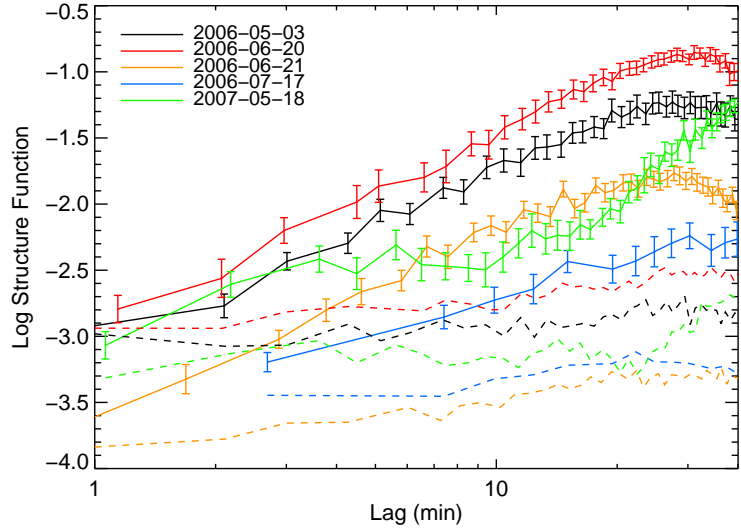}
\caption{The structure functions for five of the $K^{\prime}$ nights for Sgr A* (solid) and the comparison source, S0-17 (dashed). Error bars are calculated from $\sigma/\sqrt{N}$, where $\sigma$ is the standard deviation and $N$ the number of points in the bin. Error bars for S0-17 are of comparable sizes, but are omitted for clarity.}
\label{fig:sf}
\end{figure}

The first order structure function is often used to determine the time scale and the intrinsic variability in AGN light curves \citep[e.g.,][]{1985ApJ...296...46S,1992ApJ...396..469H,1999ASPC..159..293P}. Here, we use the structure function to determine the PSD slope for each night because it is more straightforward to calculate errors in the structure function. For a set of measurements, $s(t)$ at times $t$, the first order structure function $V(\tau)$ is defined as:
\begin{equation}
V(\tau) \equiv <[s(t+\tau) - s(t)]^{2}>
\end{equation}
Because this data set is unevenly sampled, we determined the structure function by calculating $[s(t+\tau) - s(t)]^{2}$ for all possible pairs of time lags and place them into bins with widths of one fourth of the lag times. Variable bin sizes were chosen in order to more evenly distribute the number of points in each bin, since there are many more samples at small lags than at large lags. The median lag time in each bin is assigned to be the lag time for that bin, while the average of the $V(\tau)$ values in each bin is used as the value of the structure function at that lag. The error associated with each bin is $\sigma_{bin}/\sqrt{N_{bin}}$, where $\sigma_{bin}$ is the standard deviation of the $V(\tau)$ values in the bin and $N_{bin}$ is the number of points. We do not consider lags with fewer than 5 points in that bin.  The resulting structure functions are presented in Figure \ref{fig:sf}.

\begin{figure}
\centering
\includegraphics[width=3in,angle=90]{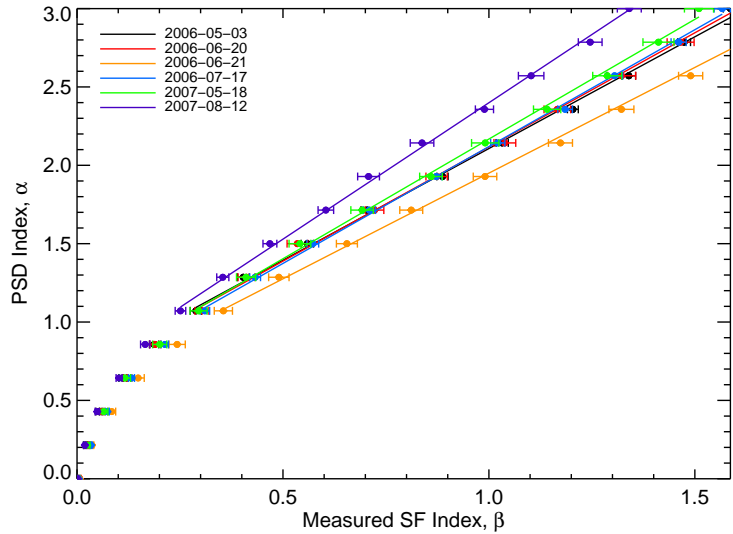}
\caption{The relationship between intrinsic PSD power index and the measured power law of the structure function as simulated for the time sampling on each night. The line fits are made for PSD power law slope $\alpha > 1.0$ in the linear region of the relationship between the intrinsic PSD power law and the structure function power law.}
\label{fig:sf_fit}
\end{figure}

Structure functions typically show a power law portion and two plateaus - one at the time scale of the noise and one plateau at lags longer than the variability time scale of the underlying physical process. The power law portion of the structure function is a measurement of the variability of the process. If the underlying process is stationary, then the logarithmic slope, $\beta$ (where $V(\tau) \propto \tau^\beta$), of this region can be related to the power index, $\alpha$ of the PSD ($P \propto f^{-\alpha}$). Because the light curves in this study have finite sampling and duration, this relationship must be determined separately for each night. To convert the measured structure function slope into the PSD slope, we measured the slope of an average of $10^3$ structure functions ($\beta$) from simulated light curves for a number of PSD slope values ($\alpha$). The relationship between the measured structure function slope and the PSD slope is linear for values of $\alpha$ between 1.0 and 3.5 (Figure \ref{fig:sf_fit}). These linear fits are used below to convert between the two slopes. We will use the structure function to compute the PSD because it is more straightforward than fitting the periodogram, given the limitations of the time sampling.

\begin{figure}
\centering
\includegraphics[width=3in,angle=90]{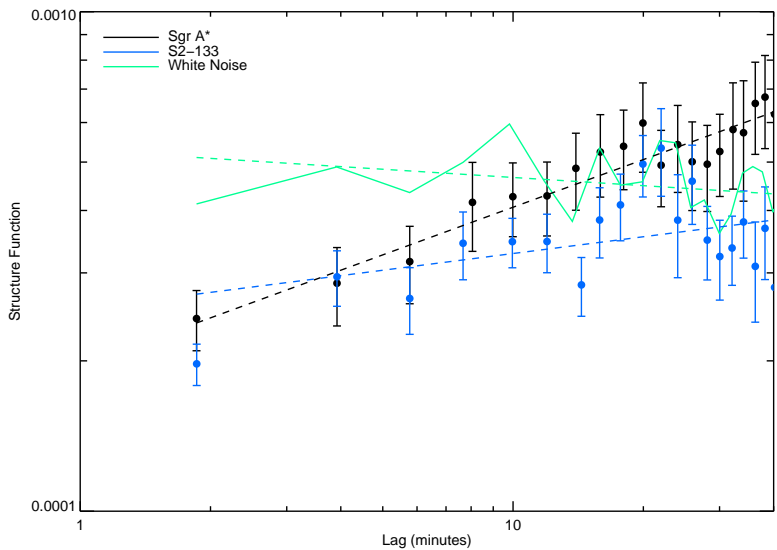}
\caption{The $K^\prime$ structure function of Sgr A* (black), S2-133 (blue), and simulations of white noise (green) for 2007 August 12..}
\label{fig:sf_aug}
\end{figure}

\begin{figure}
\centering
\includegraphics[width=3in,angle=90]{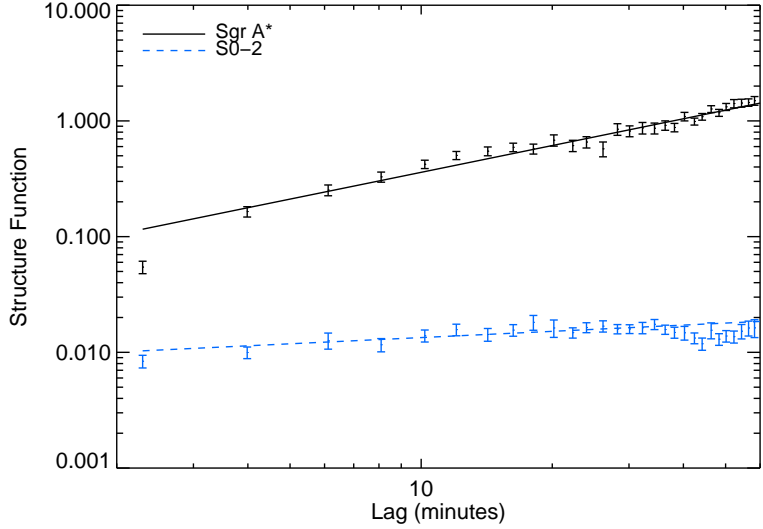}
\caption{The structure function of Sgr A* (black), S0-2 (blue) at $L^\prime$ on 2005 July 28. The best fit power law between 3 to 40 min is also plotted. The best fit power law for Sgr A* is $\beta = 0.77\pm 0.18$, corresponding to $\alpha = 1.84 \pm 0.27$ (see Section \ref{sec:sf} for details).}
\label{fig:sf_lp}
\end{figure}

The structure function for Sgr A* shows a power law portion at periods greater than about 1 minute and a flattening at periods greater than about 40 minutes. This flattening is probably not a sign of a turnover in the intrinsic variability time scale, but is rather from the poor sampling at longer periods, as shown by large variations in the value of the structure function from Monte Carlo simulations of red noise using the same time sampling. This suggests that Sgr A* is likely to be variable at all the time scales probed in this study. Fits were made to the power law region of the structure function, typically up to about half or one-third of the total time of observations for each night, and are summarized in Table \ref{table:psd}. The best fit power law index for the nights of 2006 May 3 and 2006 June 20, and 21 has $\beta = 1.4\pm0.1$ ($\alpha = 2.7\pm0.1$), between 1 to 20 minutes, consistent with the fit of $\alpha \approx 2.5$ for the slope of the combined periodogram. The other three nights have slopes ranging from  $\beta \approx 0.8$ to  $\beta \approx 0.3$. The night with the flattest slope (2007 August 12)  is also the night where the emission from Sgr A* was faint for the duration of the observation. Figure \ref{fig:sf_aug} shows that though the slope, $\beta = 0.26\pm0.04$, is flatter than on other nights, it is significantly steeper than that of S0-17 ($\beta = 0.04\pm0.04$), S2-133 ($\beta = 0.11\pm0.05$), and simulations of white noise ($\beta = -0.01$), which shows that the emission is not from a constant background source. The differences in slope between different nights may be attributable to the increase in photon noise at these faint fluxes (see Section \ref{sec:dis}). For all nights, the S0-17 structure function is also flat, with a slight rise at longer time scales (also from poor sampling). The structure function of Sgr A* for the $L^\prime$ night on 2005 July 28 has a best fit slope $\beta = 0.77\pm0.18$, corresponding to $\alpha = 1.84\pm0.27$, which is within the range of observed $K^\prime$ power law slopes (Figure \ref{fig:sf_lp}).

\subsubsection{Auto-correlation}

To compare the Sgr A* variability between each night, we computed the auto-correlation function for each observation. The auto-correlation is defined as:
\begin{equation}
AC(\tau) \equiv <[s(t+\tau)s(t)]>
\end{equation}
where $s(t)$ is the measured flux at time, $t$, and $\tau$ is the time lag. Because the light curves are unevenly sampled, we computed the auto-correlation at all possible lags and then placed the pairs in bins, similar to our method for the computing of the structure function. We find that the correlation coefficients at time scales up to 40 minutes are consistent across all nights. At time lags greater than 40 minutes, the auto-correlations appear to diverge from each other. This is mostly likely from poor sampling at those frequencies. From Monte Carlo simulations of red noise with the same time sampling as the data set, we find that the lags greater than about 40 minutes have large variations in auto-correlations coefficients; we therefore defer interpretations of lags greater than 40 minutes until more data is available to better sample these time scales.

\section{Discussion}
\label{sec:dis}

Our analysis shows that the Sgr A* near-infrared light curves observed in this study are entirely consistent with red noise with no significance evidence for a QPO signal near 20 min (or any other time scale). In addition, power is seen at the shortest time scales measured in the periodogram (2 min), well below the orbital period at the last stable circular orbit of a spinning black hole ($\sim$ 5--30 min). This indicates that the high frequency variability is not simply from Keplerian motion of a single orbiting `hot spot' producing the near-IR flux modulations \citep{2006A&A...460...15M,2007MNRAS.375..764T}. The high frequency variability must originate from small areas of the accretion flow. In addition to power at high frequencies, the light curve for 2007 May 18 shows a strong spike in flux at about 10 min with a duration of $\sim5$ min, and a rise time of only $\sim3$ min, which also suggests that regions of the accretion flow responsible for the variability can be as small as 0.4 AU. 

It is important to note that the significance threshold used in this study tests whether the periodogram power of peaks exceeds 3$\sigma$ \textit{at a given frequency}. It does not take into account that we are scanning over a large range of frequencies. If one searches enough frequencies, a peak of arbitrary power will appear at a random frequency. If a peak in the periodogram was found in this study that may indicate a periodic signal, then additional simulations should be performed to determine its statistical significance by accounting for the range of frequencies probed. This is referred to as the false alarm probability in \citet{1986ApJ...302..757H} and has been calculated analytically for the case of white noise. In the case of red noise, the additional simulations would also need to account for the fact that larger peaks are more likely at low frequencies \citep{Meyer2008}.

The timing analysis of Sgr A* in the infrared may be a way to compare the signatures of physical processes operating in the accretion flow of Sgr A* to those of AGNs. Because AGN variability studies are usually  performed on X-ray observations, the ideal comparison of the variability of Sgr A* to that of AGNs should be done at X-ray energies. However, X-ray emission from the inner accretion flow of Sgr A* can only be currently detected above the diffuse background during large flaring events. Simultaneous multi-wavelength studies of Sgr A* have shown that every detectable X-ray flare shows corresponding peaks in the near-IR, with no apparent time lag \citep{2006ApJ...644..198Y,2006A&A...450..535E}. The spectral slope is also similar at X-ray and infrared wavelengths \citep[$S_\nu \propto \nu^{-0.6}$:][]{2005ApJ...635.1095B,2006ApJ...644..198Y,2007ApJ...667..900H}, which supports a model in which the X-ray photons are the result of synchrotron self Compton scattering of lower energy photons from the population of electrons that produce synchrotron emission seen in the infrared. If the infrared and the X-ray emissions are produced by the same population of electrons, then our infrared measurements can be used as a proxy for the X-ray variability; we find that the periodograms of the infrared light curves of Sgr A* are in fact consistent with having intrinsic PSD slopes similar to those reported for AGNs at X-ray wavelengths ($\alpha \approx 1 - 3$) \citep[e.g.][]{1993ApJ...414L..85L,2001ApJ...560..659K,2003ApJ...593...96M}. Our observations are consistent with Sgr A* being variable at all time scales between $\sim$ 2 minutes to $\sim$ 1 hour with no changes in PSD slope between these time scales during the observations period on any single night. This steeply rising power law cannot be extended to arbitrarily low frequencies because the integrated power would diverge. For example, AGNs typically show either one or two breaks in their power spectrum, with recent evidence suggesting that these breaks scale with the mass and accretion rate of the AGN \citep{2006Natur.444..730M}. The lack of a break in the Sgr A* periodogram is not unusual because the breaks seen in AGNs are at time scales on the order of days or weeks \citep{2003ApJ...593...96M}. Using the relationship between the break frequency, the black hole mass, and accretion rate found by \citet{2006Natur.444..730M} for stellar mass black holes and AGNs, the predict break frequency for Sgr A* is $\sim1\times10^6$ days. However, a direct comparison to AGNs is difficult at this time because the physical scales involved are so different; the cores of AGNs in which variability is usually measured have sizes on the order of parsecs, whereas Sgr A* is unresolved in this study at a resolution of 65 mas, or $\sim$ 540 AU. The accretion luminosities of AGNs with well measured timing properties are substantial fractions of their Eddington luminosity compared to the $10^{-9}$ Eddington luminosity for Sgr A*, which may make the comparison inapplicable as well. Nevertheless, the variability properties of Sgr A* are valuable additions to the broadband spectrum since any models that can produce the observed spectrum should also be able to account for the timing characteristics. 

\begin{figure}
\centering
\includegraphics[width=3.0in,angle=90]{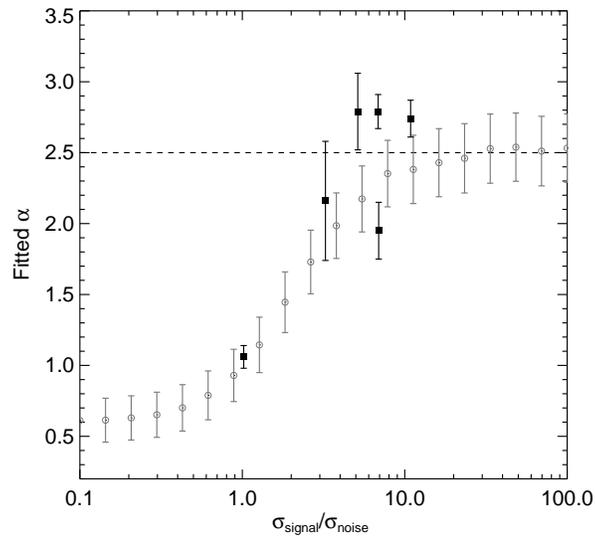}
\caption{Plot of the fitted power law slope $\alpha$ as a function of the ratio between the standard deviation of the light curve (signal) and the photon noise expected at the mean flux density for each $K^\prime$ night (solid black squares). Open grey circles shows the results of MC simulations with the same time sampling as 2007 August 12, showing that as the red noise signal becomes comparable to the measurement noise, the measured slope of the power spectrum becomes flatter than the intrinsic value of $\alpha = 2.5$ used in the simulation.}
\label{fig:alpha_snr}
\end{figure}

While it may appear from the results in Section \ref{sec:sf} that the slope of the power spectrum is changing between nights, the red noise simulations show that, as the amplitude of the red noise signal decreases, photon noise will increasingly dominate the measurement of the slope, thus biasing the measurement of the true slope of the underlying red noise process. To determine when photon noise becomes important to the slope measurements, we performed Monte Carlo simulations of light curves with time sampling from 2008 August 12 and assuming an intrinsic $\alpha = 2.5$ for various values of the ratio of the standard deviation of the red noise signal to the standard deviation of the expected measurement noise ($\sigma_{red ~noise}/\sigma_{white~ noise}$). Average values of $\alpha$ were extracted from 200 simulated light curves at each value of $\sigma_{red~noise}/\sigma_{white~noise}$ from 0.1 - 100.0 (Figure \ref{fig:alpha_snr}). We find that as $\sigma_{red~noise}/\sigma_{white~noise}$ becomes smaller, the measured value of $\alpha$ becomes flatter. The simulation showed that a measurement of $\alpha = 1.0$ can arise even from a red noise process with $\alpha=2.5$ if $\sigma_{red~noise}/\sigma_{white~noise} = 1.1$. We find that the measured value of $\alpha$ will asymptote to 2.5 when $\sigma_{red~noise}/\sigma_{white~noise} > 7.9$. This result suggests that the measurement of $\alpha = 1.06\pm0.08$ for 2007 August 12 can be explained as having the same intrinsic PSD slope, $\alpha$, as the other nights, but made flatter by a more prevalent white noise component.  Note that this is an effect of a white noise source on top of a red noise signal, which can be caused simply by the measurement noise at faint flux densities or the presence of an unresolved stellar source. Figure \ref{fig:alpha_snr} shows the relationship between the measured value of $\alpha$ and the ratio of standard deviation of the observed light curves to the expected noise level for the corresponding mean flux density ($\sigma_{signal}/\sigma_{noise}$) for all $K^\prime$ nights. The measured value of $\alpha$ appears to asymptote on nights when the red noise amplitude is stronger. This relationship is consistent with that found in the Monte Carlo simulations, where the nights with lower values of $\sigma_{signal}/\sigma_{noise}$ have flatter red noise spectra than those on the nights with greater fractional signal variance. These results show that the differences between the measured slopes for different nights may be explained by the effect of measurement noise on light curves with faint fluxes.

We also considered how a slightly time correlated source would affect the power law slope measurements by including a time correlated source with a small $\alpha$ into the simulations and varying its signal strength as the above treatment with white noise. This is motivated by the fact that there is a slight slope to the structure function of our comparison source S2-133. For example on 12 August 2007, S2-133 has a structure function power law slope of $\beta = 0.11 \pm 0.05$ corresponding to $\alpha = 0.7\pm0.2$. We find that the addition of such a background source increases the measured Sgr A* PSD power law slope $\alpha$ by a maximum of 0.1, which is about the same level as our fitting error. The effect of a slightly time correlated background source thus appears to be a source of systematic error, but one which does not significantly affect our results. 

A notable feature of the light curves in these observations is that there appears to be a lower limit, or `floor' to the light curves from each night. This lower limit is not an observational constraint, because we are able to reliably detect stars within 0\arcsec.5 of Sgr A* to K$^\prime \approx 19$ mag compared to the faintest Sgr A* observation at $K^\prime \approx 17$ mag. A question to be investigated in future studies is whether this represents a `quiescent' state of Sgr A*, characterized by lower variability at that flux level or is a faint star confused with Sgr A*. If the low flux density state is a physically distinct state, then the large flux variations would be the result of additional emission of energy above this quasi-steady state. To examine the nature of this faint state, we can look to the peak of the distribution of flux densities fainter than $\sim0.3$ mJy for each night. We find that the flux density of the peak of this distribution shifts slightly between different nights of observations. However, this shift appears only to be marginally significant given the lower photometric accuracy at the fainter flux levels. Before physical models can be applied to the faint state, a more detailed analysis of the light curves is necessary to establish whether this state exists. If the faint state of Sgr A* can be modeled as arising from the steady accretion of mass, then the slight drift of this mean flux level may be explained by a gradual change in the accretion rate. Another explanation for the drift in the mean Sgr A* flux density is that it is due to very long timescale flux variation from the red noise power law behavior of the flux as in AGNs.

It is clear that Sgr A* can now always be detected in the infrared using LGS AO and that even at low flux densities, the $K^\prime - L^\prime$ color of the source appears to be redder and have PSD slopes steeper than a star, but the measurement errors do not rule out the possibility of a faint stellar component in the near-infrared flux from the location of Sgr A*. More observations during both the faint and bright states of Sgr A* are necessary to confirm whether there truly exists a quiescent state with different timing behavior. Since Sgr A* is much redder than any stellar source, observations at $L^\prime$ would be much less affected by stellar contamination than at $K^\prime$, thus the test for a quiescent state may be more conclusive at $L^\prime$. Note that the existence of a quiescent state does not affect the timing analysis for periodic signals because the mean flux density is subtracted from a light curve to remove the zero frequency power before creating the periodogram.

While no QPO signal was detected in this study, this does not preclude the possibility for a transient periodic signal in other observations. In addition, a periodic signal with an amplitude below our detection threshold would also be difficult to detect unless the signal is consistently at the same frequency such that it would become statistically significant in a combined periodogram. More observations of Sgr A* would be helpful to investigate such low amplitude QPO signals.

\section{Conclusion}

We have obtained 7 near-infrared light curves of Sgr A* using NGS and LGS AO and through Monte Carlo simulations, test for a periodic signal in the presence of red noise with unevenly sampled data. We find no statistically significant QPO signals in any of the Sgr A* light curves observed in this study; the variability appears to be consistent with a physical process having a power spectrum index $\alpha \sim 2.5$ in the averaged periodogram, with variations between $\alpha = 1.0$ to 2.8 between different nights. A determination of the true PSD power law slope of Sgr A* will require at least double the number of light curves observed in this study so that more sophisticated spectral estimation techniques can be used \citep[e.g.][]{priestley1981,1993MNRAS.261..612P}. While more data will be necessary to reliably determine the intrinsic PSD slope, it is unnecessary to invoke a periodic signal to describe the variability of the light curves observed in this study. We have also shown the importance of preforming statistical tests for QPO signals using the entire Sgr A* light curve. Removing broad large amplitude peaks eliminates some combination of low frequency power, artificially boosting the significance of mid-range frequencies without actually removing the underlying red noise component.  

The mechanism producing the infrared variability in Sgr A* in these observations may be related to the processes occurring in AGNs or accreting black hole binaries, based on the similarity of their timing properties. Detailed model comparison between Sgr A* and other accreting black hole systems will require continuous observations over longer time scales and at faster time sampling to detect possible low and high frequency breaks, respectively, in the power spectrum. 

\acknowledgments
The authors thank the staff of the Keck observatory, especially Randy Campbell, David LeMignant, Jim Lyke, Steven McGee, and Hien Tran for all their help in obtaining the new observations and Guillaume B{\'e}langer and Matt Malkan for helpful discussions. Support for this work was provided by NSF
grant AST-0406816 and the NSF Science
\& Technology Center for AO, managed by UCSC
(AST-9876783), and the Levine-Leichtman Family Foundation.
The infrared data presented herein were obtained at the W. M. Keck Observatory, which is operated as a scientific partnership among the California Institute of Technology, the University of California and the National Aeronautics and Space Administration. The Observatory was made possible by the generous financial support of the W. M. Keck Foundation. The authors wish to recognize and acknowledge the very significant cultural role that the summit of Mauna Kea has always had within the indigenous Hawaiian community. We are most fortunate to have the opportunity to conduct observations from this mountain.

{\it Facilities:} \facility{Keck:II (NIRC2)}.

\clearpage

\begin{deluxetable}{cccccccccccccc}
\rotate
\tablecolumns{14}
\setlength{\tabcolsep}{0.03in}  % make the spacing between columns smaller
\tabletypesize{\scriptsize}  % make the font smaller so that it will fit
\tablewidth{0pc}  %make lines that go all the way to the end of the table
\tablecaption{Summary of Observations of Sgr A*}

\tablehead{\colhead{Date} & \colhead{Filter} & \colhead{Start Time} & \colhead{End Time} & \colhead{N$_{obs}$} 
& \colhead{Median Samp.} & \colhead{Duration} & \colhead{Dithered\tablenotemark{a}} & \colhead{Strehl\tablenotemark{b}} & 
\colhead{FHWM\tablenotemark{a}} & \colhead{Mean Flux\tablenotemark{c}} & \colhead{$\sigma$\tablenotemark{b}} &
\colhead{S0-17$_{phot}$\tablenotemark{d}} & \colhead{S2-133$_{phot}$\tablenotemark{d}} \\

\colhead{(UT)} & \colhead{} & \colhead{(UT)} & \colhead{UT} & \colhead{} & 
\colhead{(sec)} & \colhead{(min)} & \colhead{} & \colhead{(\%)}
& \colhead{(mas)} & \colhead{(mJy)} & \colhead{(mJy)} &
\colhead{\%} & \colhead{\%} }

\startdata
2005 July 28\tablenotemark{e} &     L$^\prime$ & 06:10:52  &  09:09:23 & 144  & 69 & 180  &  No & 70  &   81   & 4.74 &  1.7 & \nodata & \nodata  \\
2006 May 03 &     K$^\prime$ & 10:54:30  &  13:14:12 & 116  &  51 & 140  & Yes &  34   &  60     & 0.29 &  0.15 & 7 & 13\\
2006 June 20 &     K$^\prime$ & 08:59:22  &  11:17:54 & 70 & 51 & 79 & Yes &  24   &  71     & 0.24 &  0.18 & 7 & 22\\
2006 June 21 &     K$^\prime$ & 08:52:26  &  11:36:53 & 164 & 51 & 164 & Yes &  33   &  61     & 0.19 &  0.07 & 4 & 17  \\
2006 July 17\tablenotemark{f} &     K$^\prime$ & 06:45:29  &  09:54:03 & 70 & 147 & 189  & No & 35   &   59    & 0.15 &  0.05 & 5 & 22 \\
2007 May 18 &     K$^\prime$ & 11:34:10  &  13:52:39 & 77  & 51 & 84 & Yes &  36   &  59     & 0.29 &  0.14 & 7 & 10\\
2007 August 12 &     K$^\prime$ & 06:67:10  & 07:44:38  & 62   & 51 & 57 & Yes &  33   &  58     & 0.16 &  0.02 & 4 & 7 \\
\enddata
\tablenotetext{a}{When the observations are dithered, the time interval between dithers is 3 min.}
\tablenotetext{b}{Average for the night. Strehl ratios and FWHM measurements made on IRS 33N for all K$^\prime$ data and IRS 16C for L$^\prime$ data}
\tablenotetext{c}{Flux values are observed fluxes, not de-reddened.}
\tablenotetext{d}{Photometric precision for the two comparison sources S0-17 and S2-133, with mean fluxes 0.38 mJy and 0.15 mJy respectively.}
\tablenotetext{e}{NGS mode}
\tablenotetext{f}{Previously reported in \citet{2007ApJ...667..900H}}
\label{table:obs}
\end{deluxetable}

\clearpage

\begin{deluxetable}{cccc}
\tablecolumns{4}
\tablecaption{PSD fits from the structure function}
\tablewidth{0pc}
\tablehead{\colhead{Date} & \colhead{Fit Range} & \colhead{$\beta$\tablenotemark{a}} & \colhead{$\alpha$\tablenotemark{b}} \\
\colhead{(UT)} & \colhead{(min)} & \colhead{} & \colhead{} }
\startdata
2006 May 03 & 1-20 & $1.37\pm0.09$ & $2.74\pm0.13$ \\
2006 June 20 & 1-20 & $1.40\pm0.08$ & $2.79\pm0.12$ \\
2006 June 21 & 1-15 & $1.42\pm0.18$ & $2.79\pm0.27$ \\
2006 July 17 & 1-40 & $0.71\pm0.09$ & $1.80\pm0.16$ \\
2007 May 18 & 1-20 & $0.79\pm0.11$ & $1.95\pm0.20$ \\
2007 August 12 & 1-30 & $0.26\pm0.04$ & $1.06\pm0.08$ \\
\enddata
\tablenotetext{a}{Slope of the power law portion of the structure function, $V(\tau) \propto \tau^\beta$}
\tablenotetext{b}{Slope of the PSD, $P(f) \propto f^{-\alpha}$}
\label{table:psd}
\end{deluxetable}

\end{document}